\documentclass[twocolumn,showpacs,preprintnumbers,amsmath,amssymb]{revtex4}

\usepackage{amsmath,amssymb}

\usepackage{color}
\usepackage[dvips]{graphicx}
\usepackage{subfigure}

\newcommand{\lsim}{\mathrel{\mathop{\kern 0pt \rlap
  {\raise.2ex\hbox{$<$}}}
  \lower.9ex\hbox{\kern-.190em $\sim$}}}
\newcommand{\gsim}{\mathrel{\mathop{\kern 0pt \rlap
  {\raise.2ex\hbox{$>$}}}
  \lower.9ex\hbox{\kern-.190em $\sim$}}}

\newcommand{\sigman}{\sigma_{\rm scalar}^{(\rm nucleon)}}

\newcommand{\vmin}{v_{\rm min}}
\newcommand{\vesc}{v_{\rm esc}}
\newcommand{\ivmin}{{\cal I}(\vmin)}

\newcommand{\beq}{\begin{equation}}
\newcommand{\eeq}{\end{equation}}
\newcommand{\bea}{\begin{eqnarray}}
\newcommand{\ena}{\end{eqnarray}}

\begin{document}

\preprint{DFTT 24/2005}

\title{Do current WIMP direct measurements constrain light relic neutralinos?}



%
\author{A. Bottino, F. Donato, N. Fornengo}
\affiliation{Dipartimento di Fisica Teorica, Universit\`a di Torino \\
Istituto Nazionale di Fisica Nucleare, via P. Giuria 1, I--10125 Torino, Italy}

\author{S. Scopel}
\affiliation{School of Physics,
Korea Institute for Advanced Study\\ 207-43 Cheongryangri-dong,
Dongdaemun-gu, Seoul 130-012, Korea}

\date{\today}

\begin{abstract}
New upper bounds on direct detection rates have recently been
presented by a number of experimental collaborations working on
searches for WIMPs. In this paper we analyze how the constraints on
relic neutralinos which can be derived from these results is affected
by the uncertainties in the distribution function of WIMPs in the
halo. Various different categories of velocity distribution functions
are considered, and the ensuing implications for supersymmetric
configurations derived. We conservatively conclude that current
experimental data do not constrain neutralinos of small mass
(below 50 GeV).
\end{abstract}


\maketitle


\section{Introduction}
\label{sect:intro}

In Refs. \cite{lowneu,lowmassdir,lowind,lowpbar} we have discussed the
cosmological properties of light neutralinos ({\it i.e.} neutralinos
with a mass in the range 6 GeV $\lsim m_{\chi} \lsim$ 50 GeV), which
originate in supersymmetric schemes where gaugino-mass unification is
not assumed.  Actually, the most remarkable features occur for
neutralinos in the mass range: 7 GeV $\lsim m_{\chi} \lsim$ 25 GeV.
Namely, for relic neutralinos with these masses, direct and indirect
detection rates are considerably high, and at the level of present
experimental sensitivities. Furthermore, the range of the predicted
values for the rates is quite narrow, at variance with what happens
for neutralinos of higher masses, where the expected rates are spread
over decades.

The properties of light neutralinos with respect to WIMP direct
measurements were analyzed in Refs. \cite{lowmassdir,lowind}.  Since
then, some new results and/or analyses of previous data from
experiments of WIMP direct searches have appeared
\cite{cresst,edelweiss,zeplin,cdms}. In the present paper, we examine
whether these new data put some constraints on the relic neutralinos
of light masses.

Let us recall that the differential event rate $dR/dE_R$ ($E_R$ being
the nuclear recoil energy) measured in WIMP direct searches is a
convolution of the WIMP-nucleus cross section with the WIMP
phase-space distribution function of WIMPs, evaluated at the Earth
location. By assuming that: i) in this phase-space distribution
function, spatial and velocity dependence factorize, and 2) coherent
interactions dominate over incoherent ones in the WIMP-nucleus
scattering (which is usually the case for relic neutralinos), one
recovers the expression:

\begin{equation}
\frac{dR}{dE_R} = N_T \frac{\rho_0}{m_\chi}\frac{m_N}{2 \mu_1^2} 
A^2 \xi \sigma_{\rm scalar}^{(\rm nucleon)} F^2(E_R)\, {\cal I}(v_{\rm min})\, ,
\label{eq:rate}
\end{equation}
where:
\begin{equation}
{\cal I}(v_{\rm min}) = \int_{w \geq v_{\rm min}} d^3 w \;\;\frac{f_{\rm ES}(\vec w)}{w}\, .
\label{eq:ivmin}
\end{equation}
In the previous formulae, notations are: $N_T$ is the number of target
nuclei per unit mass, $m_\chi$ is the WIMP mass, $m_N$ is the nucleus
mass, $\mu_1$ is the WIMP--{\em nucleon} reduced mass, $A$ the nuclear
mass number, $\sigman$ is the WIMP--nucleon coherent cross section,
$F(E_R)$ is the nuclear form factor, $\xi$ is the fraction of the mass
density of the WIMP in terms of the total local density for
non-baryonic dark matter $\rho_0$ ({\it i.e.}: $\xi = \rho_W/\rho_0$),
$f_{\rm ES}(\vec w)$ and $\vec w$ denote the velocity distribution
function (DF) and WIMP velocity in the Earth frame, respectively
($w=|\vec w|$). It is natural to define the velocity distribution
function in the Galactic rest frame $f(\vec{v})$, where $\vec{v} =
\vec{w} + \vec{v}_{\oplus}$, $\vec{v}_{\oplus}$ being the Earth
velocity in the Galactic rest frame. The Earth frame velocity DF is
then obtained by means of the transformation: $f_{\rm ES}(\vec
w)=f(\vec w + \vec v_\oplus)$. It is implicitly understood that the
velocity DF $f(\vec v)$ is truncated at a maximal escape velocity
$v_{\rm esc}$, since the gravitational field of the Galaxy cannot
bound arbitrarily fast WIMPs. The value we adopt here is: $v_{\rm esc}
= 650$ km sec$^{-1}$ \cite{escape}, although we will
comment on the effect of a lower value, which we will set at $v_{\rm
esc} = 450$ km sec$^{-1}$ \cite{escape}. Finally, the quantity
$v_{\rm min}$ appearing in Eq. (\ref{eq:ivmin}) defines the minimal
Earth--frame WIMP velocity which contributes to a given recoil energy
$E_R$:
\begin{equation}
v_{\rm min} = [m_N E_R/(2 \mu_A^2)]^{1/2}\, ,
\label{eq:vmin}
\end{equation}
where $\mu_A$ is the WIMP--{\em nucleus} reduced mass.

Eqs. (\ref{eq:rate}) and (\ref{eq:ivmin}) are the basis for deriving
information on the quantity $\xi \sigman$ from the measurements on the
differential rate $dR/dE_R$. However, this procedure implies the use
of a specific WIMP distribution function, which determines both the
value of the local dark matter density $\rho_0$ and the shape of the
velocity DF $f(\vec v)$.

In the present paper we first discuss how upper bounds on $\xi
\sigman$, derived from experimental upper limits on $dR/dE_R$, depend
on the large uncertainties affecting the WIMP distribution
functions. We then discuss what is the relevance of these upper bounds
on $\xi \sigman$ for light relic neutralinos.

\section{WIMP distribution functions}
\begin{table}
\begin{ruledtabular}
\begin{tabular}{|lc|}
\multicolumn{2}{|l|}{{\bf A: Spherical $\bf \rho_{DM}$,  isotropic
  velocity dispersion}} \\ 
\hline
A0 & {\rm ~Isothermal sphere} \\
A1 & {\rm ~Evans' logarithmic} \cite{evans_log} \\
A2 & {\rm ~Evans' power--law} \cite{evans_pl} \\
A5 & {\rm ~NFW} \cite{nfw} \\
~&~\\ 
\hline 
\multicolumn{2}{|l|}{{\bf B: Spherical $\bf \rho_{DM}$, non--isotropic
  velocity dispersion}} \\ 
\hline 
B1 & {\rm ~Evans' logarithmic} \cite{evans_log} \\
B2 & {\rm ~Evans' power--law} \cite{evans_pl} \\
B5 & {\rm ~NFW} \cite{nfw} \\
~&~\\ 
\hline 
\multicolumn{2}{|l|}{{\bf C: Axisymmetric $\bf \rho_{DM}$}}  \\ 
\hline
C2 & {\rm ~Evans' logarithmic}  \\
C3 & {\rm ~Evans' power--law}  \\
~&~\\ 
\hline 
\multicolumn{2}{|l|}{{\bf D: Triaxial $\bf \rho_{DM}$ \cite{triaxial}}}  \\ 
\hline
D1 & {\rm ~Earth on major axis, radial anisotropy}  \\
\end{tabular}
\end{ruledtabular}
\caption{\label{tab:models} Summary of the galactic halo models
  considered in our analysis.  The label shown in the first
  column is used throughout the text to indicate each model in a
  unique way and corresponds to the classification introduced in
  Ref. \cite{bcfs}. For details on the models and proper definitions,
  see Ref. \cite{bcfs}.}
\end{table}

\begin{table*}
\begin{ruledtabular}
\begin{tabular}{cccc}
 & {$v_0 =170$ km sec$^{-1}$} &  {$v_0 =220$ km sec$^{-1}$} &  {$v_0 =270$ km sec$^{-1}$} \\ 
\hline
Model &  $\rho_0 ~({\rm GeV\,cm^{-3})}$  & $\rho_0 ~({\rm GeV\,cm^{-3})}$ &  $\rho_0 ~({\rm GeV\,cm^{-3})}$ \\ 
\hline 
  A0           & 0.18 & 0.30 & 0.71 \\
  A1 ,  B1     & 0.20 & 0.34 & 1.07 \\
  A2 ,  B2     & 0.24 & 0.41 & 1.33 \\
  A5 ,  B5     & 0.20 & 0.33 & 1.11 \\
  C2           & 0.67 & 1.11 & 1.68 \\
  C3           & 0.66 & 1.10 & 1.66 \\
  D1           & 0.50 & 0.84 & 1.27 \\
\end{tabular}
\end{ruledtabular}
\caption{\label{tab:intervals} Values of the dark matter local density
  $\rho_0$ corresponding to the three different values of the local
  rotational velocity $v_0$ and obtained from the constraints on the
  amount of non--halo component and on the flatness of the galactic
  rotational curve, for the different halo models of Table
  \ref{tab:models}. For the models of class A and B, the values of
  $\rho_0$ are the minimal ones for $v_0 =170$ km sec$^{-1}$ and $v_0
  =220$ km sec$^{-1}$ ({\em i.e.}  corresponding to a minimal halo
  contribution), while for $v_0 =270$ km sec$^{-1}$ the values of
  $\rho_0$ are the maximal ones (referring to a maximal halo).  For
  models of class C and D, the value of $\rho_0$ is always the maximal
  one. The axisymmetric models of class C are not affected by the
  inclusion of a co--rotation or counter--rotation effect.}
\end{table*}

\begin{figure} \centering
\vspace{-40pt}\includegraphics[width=1.1\columnwidth]{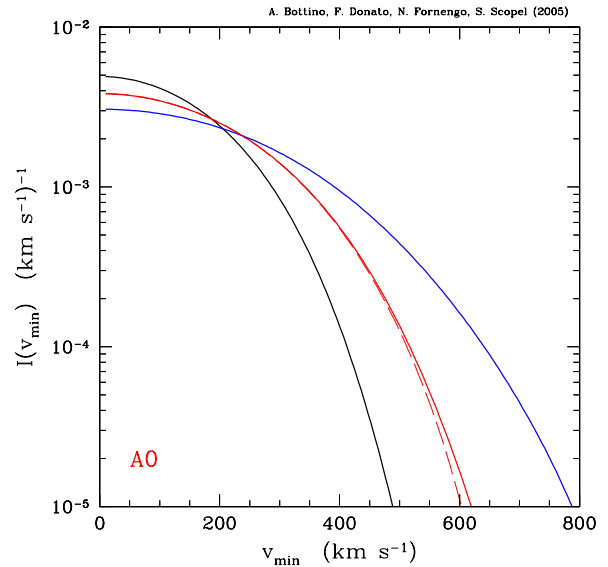}
\vspace{-30pt}\caption{\label{fig:ivmin_isothermal} Function $\ivmin$
 for an isothermal sphere (model A0 of Table \ref{tab:models}) and
 $\vesc=650$ km sec$^{-1}$. The solid curves (from top to bottom, as
 seen on the extreme left of the plot) refer to the three values of
 $v_0$ (and ensuing $\rho_0$) given in Table \ref{tab:intervals}:
 $v_0=170$ km sec$^{-1}$ (top), $v_0=220$ km sec$^{-1}$ (medium),
 $v_0=270$ km sec$^{-1}$ (bottom).  The dashed line shows the
 modification of the median isothermal case when $\vesc=450$ km
 sec$^{-1}$.}
\end{figure}

\begin{figure} \centering
\vspace{-40pt}\includegraphics[width=1.1\columnwidth]{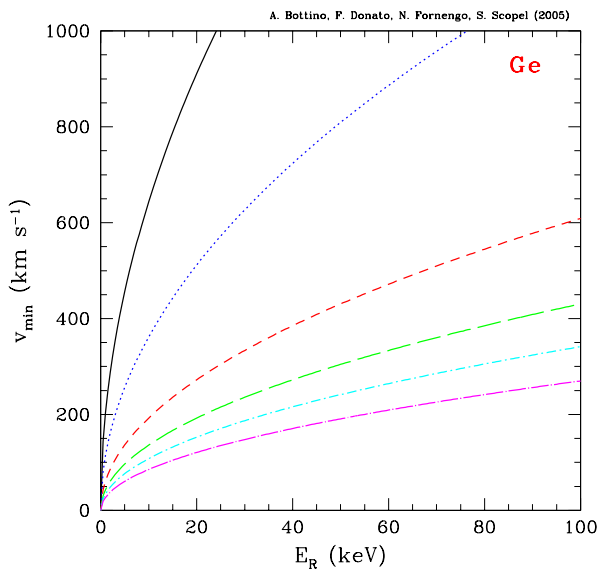}
\vspace{-30pt}\caption{\label{fig:Ge_vmin}Values of $\vmin$ as a
function of the nuclear recoil energy $E_R$, for a Ge detector.  The
different curves refer to WIMP masses of: 10, 20, 50, 100, 200 GeV and
1 TeV, from top to bottom.}
\end{figure}

\begin{figure*} \centering
\vspace{-40pt}\includegraphics[width=2.2\columnwidth]{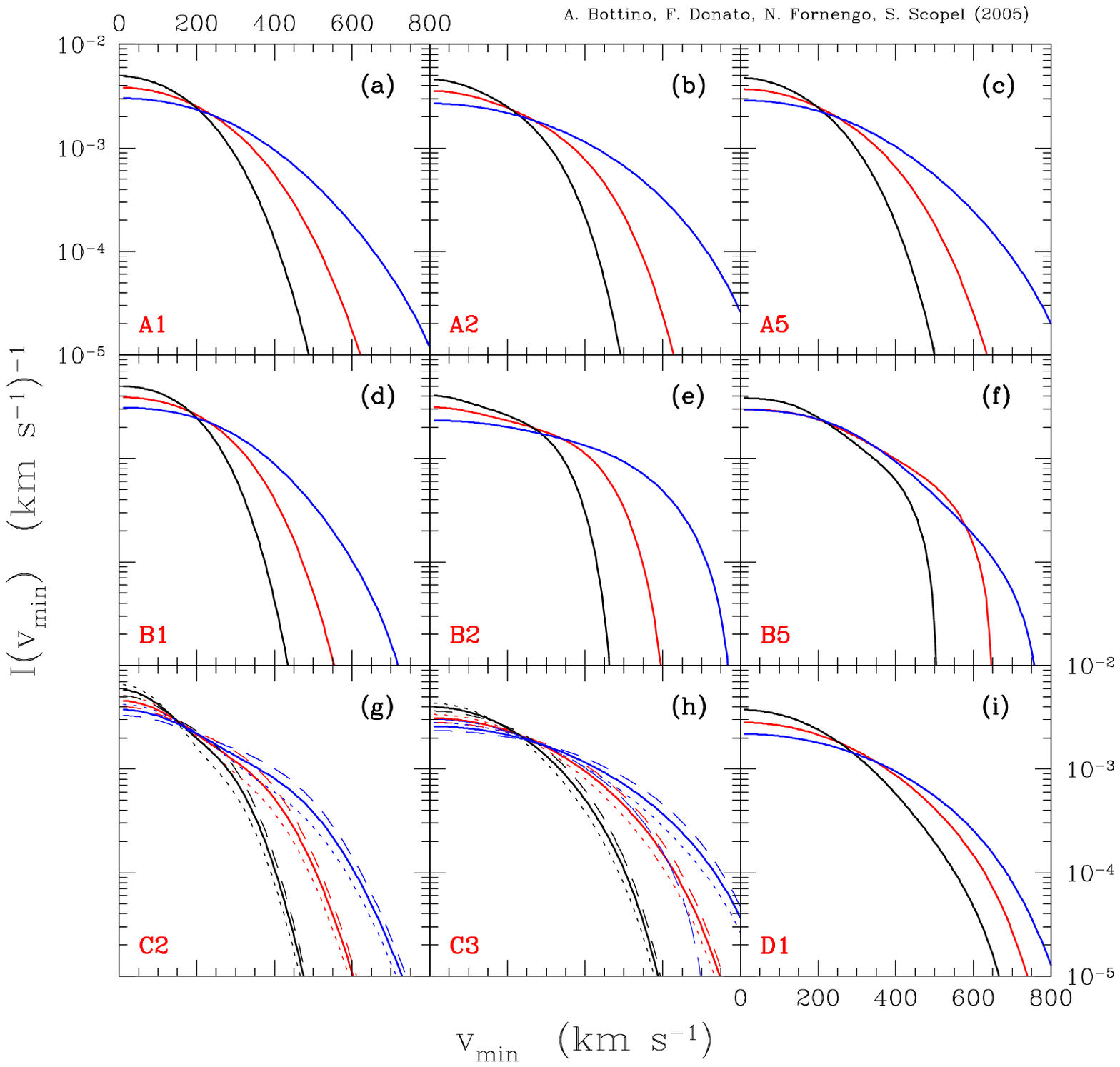}
\vspace{-60pt}\caption{\label{fig:ivmin_distribs}Function $\ivmin$ for
 all the galactic model of Table \ref{tab:models}, other than the
 isothermal sphere, for $\vesc=650$ km sec$^{-1}$. The label which
 identifies the model is written in the bottom--left corner of each
 panel. Notations are as in Fig. \ref{fig:ivmin_isothermal}. In panels
 (g) and (h), which correspond to axisymmetric models, the dotted and
 dashed lines refer to maximal galactic co--rotation and
 counter--rotation, respectively. In panel (h), the
 long--dashed line shows the modification of the $v_0=270$ km
 sec$^{-1}$ case when $\vesc=450$ km sec$^{-1}$.}
\end{figure*}

In our analysis we consider a subset of the large sample of galactic
halo models which were studied in detail in
Ref. \cite{bcfs}. Following Ref. \cite{bcfs}, we classify the DFs into
four categories, depending on the symmetry properties of the matter
density (or the corresponding gravitational potential) and of the
velocity dependence: A) spherically symmetric matter density
$\rho_{\rm DM}$ with isotropic velocity dispersion, B) spherically
symmetric matter density with non--isotropic velocity dispersion, C)
axisymmetric models, D) triaxial models \cite{note,others}

For each category, different specific models are identified. The
models considered in the present analysis are listed in Table
\ref{tab:models}. For a thorough definition of the different models
and of the values of their intrinsic parameters, and for a detailed
description of theoretical technicalities, we refer to Ref.
\cite{bcfs}. Here we just remind that for each model we calculate,
either analytically (when possible) or numerically, the velocity DF
which accompanies a given matter density distribution. For the
spherically symmetric and isotropic models of class A, the velocity DF
is obtained by solving the Eddington equation \cite{bcfs,binney}. For
the spherically symmetric and non--isotropic models of class B we
assume the anisotropy to be described in terms of the Osipkov--Merrit
parameter $\beta$ \cite{bcfs,binney,osipkov_merrit} which defines the
degree of anisotropy (we fix $\beta=0.4$): in this case, the velocity
DF can be obtained by a generalization of the Eddington method
\cite{bcfs,binney}. Axisymmetric models of class C allow the presence
of a definite angular momentum. We choose them as a direct
generalization of some of the models of class A: for these models,
analytical solutions to the relevant generalized Eddington equation
may be found. In this case, we also allow for a maximal co--rotation
or counter--rotation of the galactic halo. Finally, class D presents a
specific triaxial model.

Each halo model is constrained by a number of observational inputs
\cite{bcfs}: i) properties of the galactic rotational curve, namely
the range of the allowed values for the local rotational velocity,
$170~\mbox{km sec$^{-1}$}\leq v_0 \leq 270~\mbox{km sec$^{-1}$}$
\cite{kochanek,cepheids}, and the amount of flatness of the rotational
curve at large distances from the galactic center, and ii) the maximal
amount of non--halo components in the Galaxy, $M_{\rm vis}$ ({\em
i.e.} the disk, the bulge, etc.). These constraints determine first of
all the value of the local dark matter density $\rho_0$, which is a
relevant parameter in the direct detection rate. Depending on whether
one allows for a maximal halo ({\it i.e.} the contribution of the
non-halo components is minimized) or, the other way around, the
contribution of the halo to the rotational curve is minimized, the
value of $\rho_0$ is either increased or reduced, respectively. This
is discussed in detail in Ref. \cite{bcfs} and is manifest also in
Table \ref{tab:intervals}, where values of $\rho_0$ for each halo
model and for the three representative values of $v_0$ are shown. The
difference in the values of $\rho_0$ depending on the assumption of a
minimal or a maximal halo contribution is given in Ref. \cite{bcfs}. For
instance, for the isothermal sphere $\rho_0$ falls in the range
$(0.18,0.28)~\mbox{GeV cm$^{-3}$}$ for $v_0 = 170~\mbox{km
sec$^{-1}$}$, $(0.30,0.47)~\mbox{GeV cm$^{-3}$}$ for $v_0 =
220~\mbox{km sec$^{-1}$}$, and $(0.45,0.71)~\mbox{GeV cm$^{-3}$}$ for
$v_0 = 270~\mbox{km sec$^{-1}$}$. We therefore expect a significant
variability in the extracted upper limits in direct detection
experiments also for the simple and generally used isothermal sphere
model. We notice that the standard reference choice of
$\rho_0=0.30~\mbox{GeV cm$^{-3}$}$ for $v_0 = 220~\mbox{km
sec$^{-1}$}$ refers to the case of a minimal halo.

The choice we make here for the values of $\rho_0$ associated to each
representative value of $v_0$ is the following: for $v_0=170~\mbox{km
sec$^{-1}$}$ (which correspond to its 95\% C.L. lower bound) we adopt
the case of a minimal halo, in order to determine the set of less
constraining upper--limits on $\xi\sigman$, as is clear from
Eq. (\ref{eq:rate}).  For $v_0=270~\mbox{km sec$^{-1}$}$ (which
correspond to its 95\% C.L. upper bound) we instead adopt a maximal
halo: in this case we will determine the most constraining
upper--limits on $\xi\sigman$. In the case of the central (and
reference) value $v_0 = 220~\mbox{km sec$^{-1}$}$, we adopt a minimal
halo, which reproduces the standard choice $\rho_0=0.30~\mbox{GeV
cm$^{-3}$}$ for the isothermal sphere. These considerations apply to
all models of class A and B. For the models of class C and D, for
which we can rely only on analytical solutions for the velocity DF, we
are forced to use always the case of a maximal halo: in fact,
analytical solutions of class C and D are actually obtained only for a
maximal halo contribution \cite{bcfs}.

Let us turn now to the direct effect of the velocity DF on the
detection rate, which is studied here in terms of the relevant
function $\ivmin$ of
Eq. (\ref{eq:ivmin}). Fig. \ref{fig:ivmin_isothermal} shows $\ivmin$
for the isothermal halo and for the three values of $v_0$ listed in
Table \ref{tab:intervals}. We see that for low values of $\vmin$ the
larger contribution to the detection rate occurs when $v_0$ is
smaller, since for smaller rotational velocities the velocity
dispersion of the isothermal Maxwellian distribution is also smaller,
and in turn this enhances the average inverse velocity, which is
related to the definition of $\ivmin$: $\langle 1/w \rangle = {\cal
I}(0)$. This can be analytically understood by remembering that for a
pure isothermal sphere and a maximal halo the velocity distribution
function is just an isotropic Maxwellian with velocity dispersion
given by $v_0$:
\begin{equation}
f_A(v) = [\pi v_0^2]^{-3/2} \exp(-v^2/v_0^2)\, ;
\label{eq:fA} 
\end{equation}
in this case the function $\ivmin$ reads (in the limit $\vesc
\rightarrow \infty$) \cite{fds}:
\begin{equation}
\ivmin = \frac{1}{2\eta v_0}[{\rm erf}(x_{\rm min}+\eta) - {\rm erf}(x_{\rm min}-\eta)]\, ,
\label{eq:ivminA} 
\end{equation}
where $x_{\rm min} = v_{\rm min}/v_0$ and $\eta = v_\oplus/v_0$. From
Eq.(\ref{eq:ivminA}) we see that for small values of $\vmin$ the
larger $\ivmin$ occurs for smaller $v_0$ because of the inverse 
law dependence.

On the contrary, for large values of $\vmin$, the almost--exponential
tail in $\ivmin$ is more severe when $v_0$ is small, and therefore the
behaviour of $\ivmin$ with respect to $v_0$ is the opposite. This again is
understood from the simple expression of Eq.(\ref{eq:ivminA}). The
regime we are considering ($\vmin \gsim v_\oplus$) asymptotically can
be studied as the limit $\eta \rightarrow 0$ in Eq. (\ref{eq:ivminA}),
which gives:
\begin{equation}
\ivmin = 2[\pi v_0^2]^{-1/2} \exp(-\vmin^2/v_0^2)\, .
\label{eq:ivminAlim} 
\end{equation}
This shows the discussed behaviour as a function of $v_0$. The tail,
due to the presence of a non vanishing $\eta$ in Eq. (\ref{eq:ivminA}),
is less severe than the one in Eq. (\ref{eq:ivminAlim}) but
nevertheless it follows the same behaviour. 

Also the value of the escape velocity is relevant in the large $\vmin$
tail of the function $\ivmin$.  The results presented so far in
Fig. \ref{fig:ivmin_isothermal} are obtained for a value of the escape
velocity (in the galactic frame) $\vesc= 650~\mbox{km
sec$^{-1}$}$. However, values as low as $\vesc= 450~\mbox{km
sec$^{-1}$}$ have also been considered \cite{escape}. A lower escape
velocity implies a cut in the high $\vmin$ tail of $\ivmin$
\cite{fds}. This effect is shown in Fig. \ref{fig:ivmin_isothermal}
for the central case $v_0=220~\mbox{km sec$^{-1}$}$.

The discussion on the behaviour of $\ivmin$ has direct impact on the
direct detection rate, since $\vmin$ is directly related to the recoil
energy. Eq. (\ref{eq:vmin}) implies that very light WIMPs can produce
recoil energies in the tens of keV range only if they possess large
velocities. In this case, the detection rate for such light WIMPs
will be mostly determined by the almost--exponential tail in the
function $\ivmin$ discussed above.  On the contrary, heavy WIMPs can
produce recoil in the same tens of keV range by possessing much lower
velocities: they will be therefore more sensitive also to the low
$\vmin$ part of the function $\ivmin$.  The quantitative connection
between $v_{\rm min}$ and $E_R$ for a Ge nucleus and different WIMP
masses is given in Fig. \ref{fig:Ge_vmin}.

Finally, Fig. \ref{fig:ivmin_distribs} shows the function $\ivmin$ for
all the other halo models listed in Table \ref{tab:models}. As for the
symmetric and isotropic models, we see that in the case of a
power--law behaviour of the gravitational potential (model A2) or for
the NFW density profile (model A5) the large $\vmin$ tail is less
suppressed, mainly for low $v_0$: this comes along with a larger
predicted detection rate and it will translate into a more
constraining upper limit for WIMPs lighter than a few tens of GeV. In
the case of anisotropic models, we notice that the most direct
anisotropic generalization of the isothermal sphere, which is a cored
spherical distribution with anisotropic velocity dispersion (model B1)
is the one which is more suppressed at large $\vmin$: this has the
effect of reducing the sensitivity of the detector to light WIMPs. On
the contrary, the axisymmetric model with a power--law gravitational
potential C3 is the one with the highest tail in the function
$\ivmin$, and therefore more sensitive in constraining light WIMPs.
We also notice that for this type of models, which possess an enhaced
$\vmin$ tail, the effect of a lower escape velocity is more dramatic:
panel (h) of Fig. \ref{fig:ivmin_distribs} shows the sizeable
reduction of $\ivmin$ at large $\vmin$ for an escape velocity
$\vesc=450$ km sec$^{-1}$ in the case of model C3 and $v_0=270$ km
sec$^{-1}$.

The local matter density values $\rho_0$ of Table \ref{tab:intervals}
and the results of Fig. \ref{fig:ivmin_isothermal} and
Fig. \ref{fig:ivmin_distribs} are the key elements which will be used
in the next section to determine upper limits on the WIMP--nucleon
scattering cross sections.

\section{Results and conclusions}
\begin{figure} \centering
\vspace{-40pt}\includegraphics[width=1.1\columnwidth]{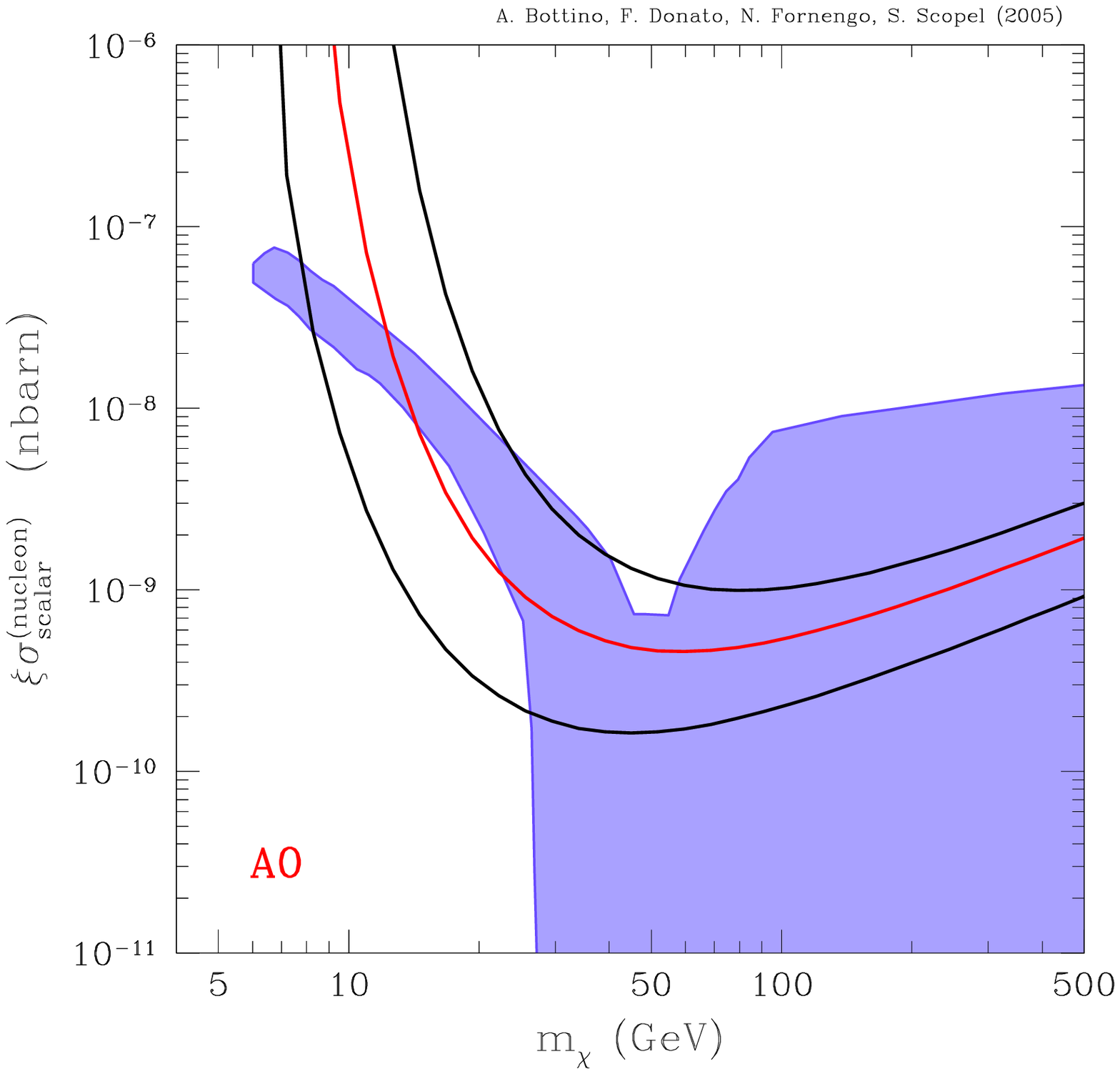}
\vspace{-30pt}\caption{\label{fig:limits_isothermal} The solid lines
show the upper limit on the quantity $\xi \sigman$ as a function of
the WIMP mass $m_\chi$ for the CDMS detector and for an isothermal
sphere (model A0 of Table \ref{tab:models}). The curves refer to the
three values of $v_0$ (and corresponding $\rho_0$) given in Table
\ref{tab:intervals}: $v_0=170$ km sec$^{-1}$ (top), $v_0=220$ km
sec$^{-1}$ (medium), $v_0=270$ km sec$^{-1}$ (bottom). The colored
region shows the values of $\xi \sigman$ for neutralino dark matter,
obtained in a scan of the minimal supersymmetric model defined in
Refs. \cite{lowneu,lowmassdir,lowind,lowpbar}. The funnel for low
neutralino masses (below 50 GeV) corresponds to supersymmetric models
without gaugino--mass unification.}
\end{figure}

\begin{figure} \centering
\vspace{-40pt}\includegraphics[width=1.1\columnwidth]{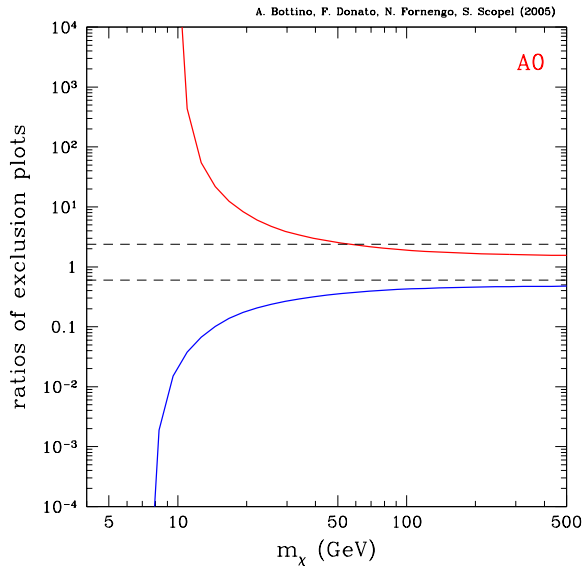}
\vspace{-30pt}\caption{\label{fig:ratio} Ratio of the upper limits of
Fig. \ref{fig:limits_isothermal}, obtained for an isothermal sphere
(model A0). The upper curve is the ratio between the upper and the
central curves in Fig. \ref{fig:limits_isothermal}. The lower curve is
the ratio between the lower and central curves}
\end{figure}

\begin{figure*} \centering
\vspace{-40pt}\includegraphics[width=2.2\columnwidth]{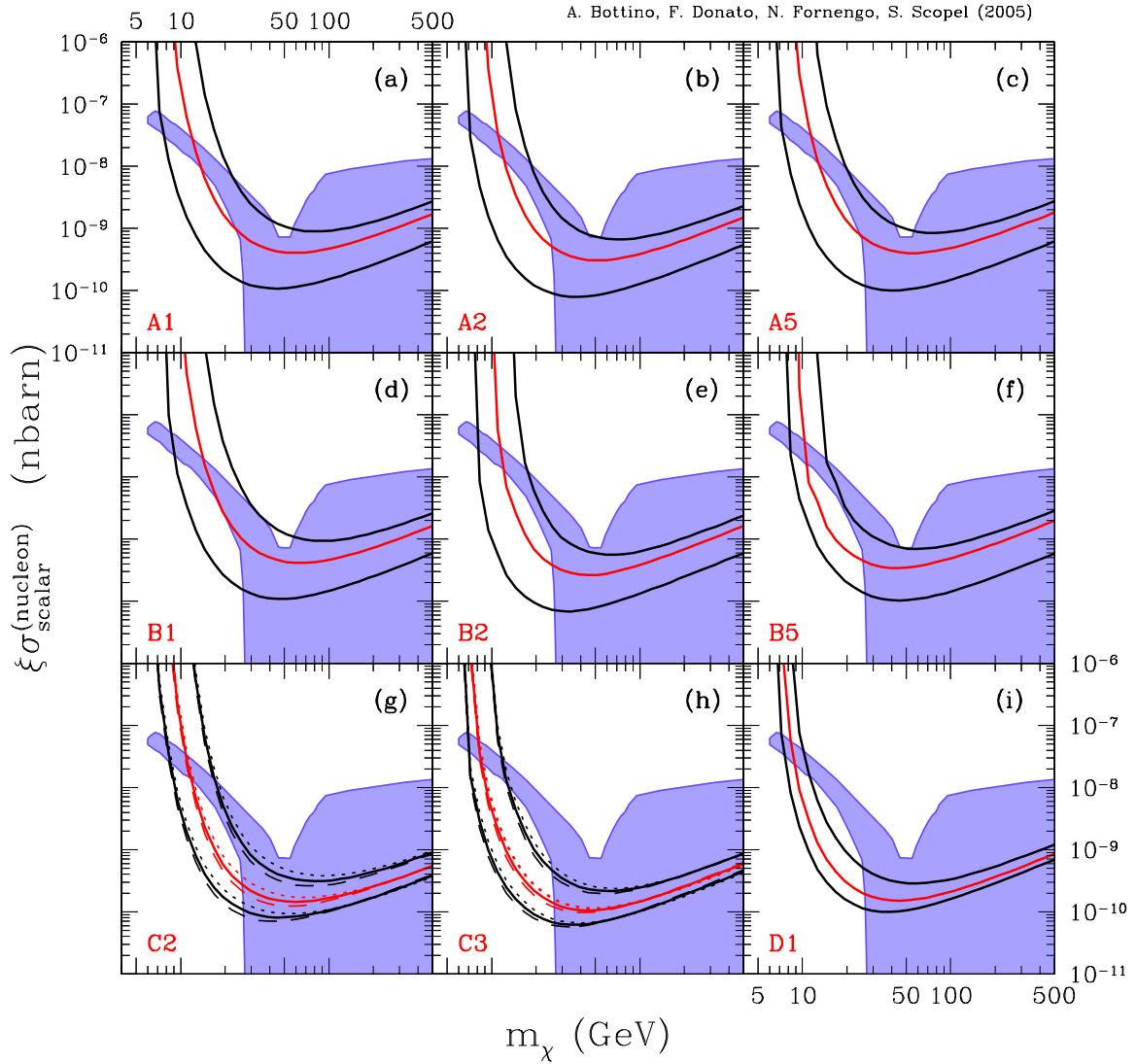}
\vspace{-60pt}\caption{\label{fig:limits_distribs}The solid lines show
the upper limit on the quantity $\xi \sigman$ as a function of the
WIMP mass $m_\chi$ for the CDMS detector and for the different
galactic models of Table \ref{tab:models}, other than the isothermal
sphere. The label which identifies the model is written in the
bottom--left corner of each panel. Notations are as in
Fig. \ref{fig:limits_isothermal}. In panels (g) and (h), which
correspond to axisymmetric models, the dotted and dashed lines refer
to maximal galactic co--rotation and counter--rotation, respectively.}
\end{figure*}

\begin{figure} \centering
\vspace{-40pt}\includegraphics[width=1.1\columnwidth]{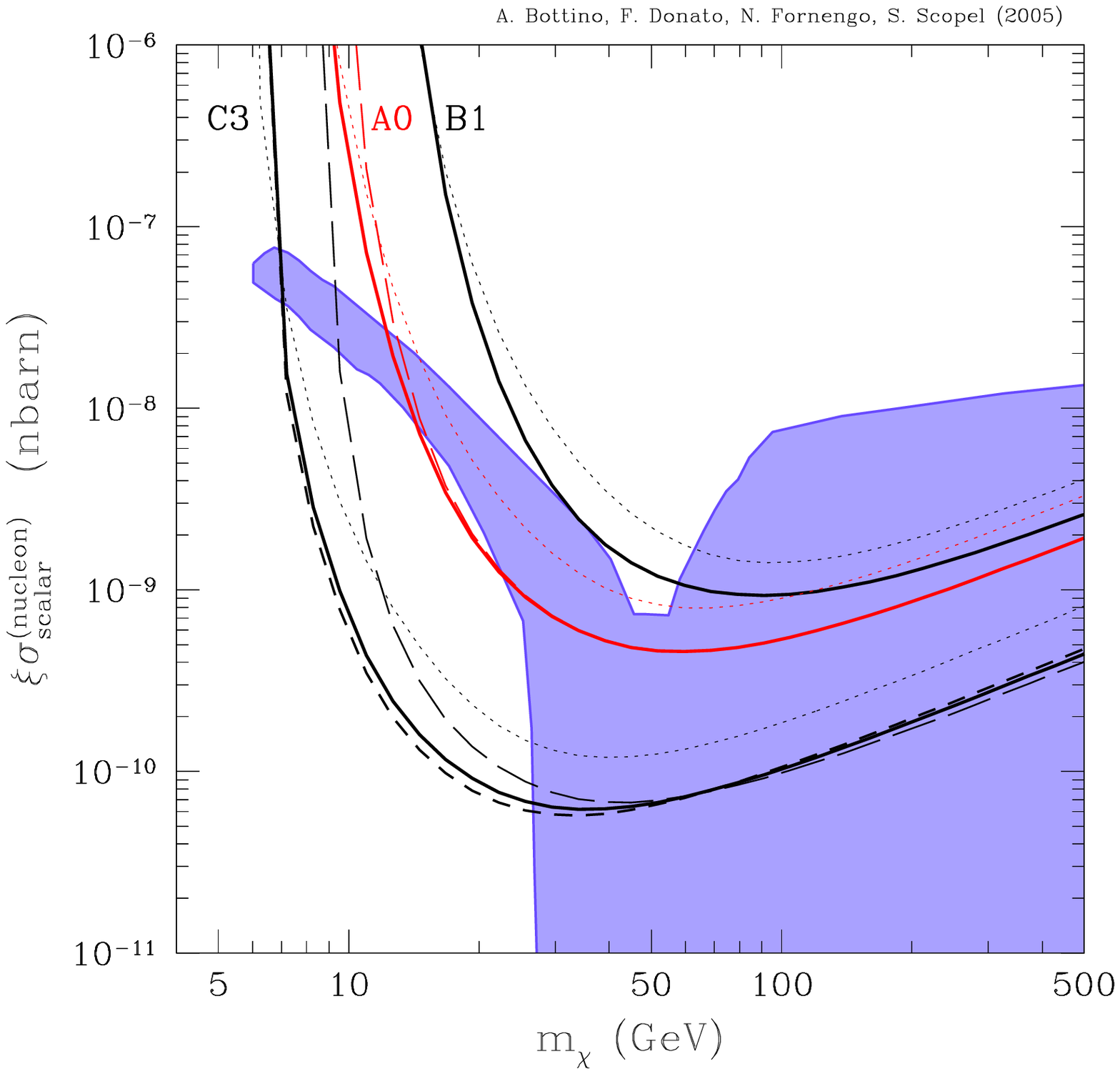}
\vspace{-30pt}\caption{\label{fig:limits_extreme}The solid lines show
the summary of our analysis on the upper limit on the quantity $\xi
\sigman$ as a function of the WIMP mass $m_\chi$ for the CDMS detector
and for $\vesc=650$ km sec$^{-1}$. The median line refers to the
standard isothermal sphere with $v_0 =220$ km sec$^{-1}$ and $\rho_0
=0.3 ~{\rm GeV\,cm^{-3}}$ (model A0). The upper and lower curves show
the two extremes obtained in the analysis and refer to model B1 with
$v_0 =170$ km sec$^{-1}$ (upper solid line) and model C3 with $v_0
=270$ km sec$^{-1}$ (lower solid line). The dashed line refers to
model C3 with maximal counter--rotation of the galactic halo. The
dotted lines show the ZEPLIN I limits obtained for the same galactic
models. The long--dashed lines show the upper limits for CDMS in the
case of a lower escape velocity $\vesc=450$ km sec$^{-1}$: the upper
line refers to model A0, the lower one to model C3. For model B1, the
limit coincides with the corresponding solid line.  The colored region
shows the values of $\xi \sigman$ for neutralino dark matter, obtained
in a scan of the minimal supersymmetric model defined in
Refs. \cite{lowneu,lowmassdir,lowind,lowpbar}. The funnel for low
neutralino masses (below 50 GeV) corresponds to supersymmetric models
without gaugino--mass unification.}
\end{figure}

In Refs. \cite{cresst,edelweiss,zeplin,cdms} upper limits on $\xi
\sigman$ are obtained by using a standard isothermal distribution
function with density parameter $\rho_0=0.3$ GeV cm$^{-3}$ and
$v_0=220$ km sec$^{-1}$. Here, we analyze this class of experimental
data employing the sample of distribution functions discussed in the
previous section. Our analysis and discussion is performed in terms of
the experimental data of Ref. \cite{cdms} (CDMS), since these turn out
to be the most constraining ones. For very light WIMPs, with masses
below 10 GeV, also the results of Ref. \cite{zeplin} (ZEPLIN) may play
a role, and we will add them in our final discussion.

In order to obtain the exclusion--plot from the CDMS data, we extract
the neutralino--nucleon cross section that yields a number of events
compatible with zero between 10 keV and 64 keV with an effective
exposure $(MT)_{\rm eff} = 19.4$ kg day, which corresponds to the
effective exposure for $m_\chi=60$ GeV quoted in Ref. \cite{cdms}.  By
considering a Poissonian fluctuation of the expected rate, we assume
the upper bound of 2.3 events at 90\% C.L. In the calculation of the
expected rate, we use Helm nuclear form factors and a bolometric
quenching factor equal to 1, as quoted by the experimental
Collaboration \cite{cdms}. Our procedure for extracting the
exclusion--plot is less refined than the one adopted in
Ref. \cite{cdms}, since it neglects the dependence of $(MT)_{\rm
eff}$ on the WIMP mass and does not take into account the part of the
CDMS spectrum between 64 and 100 keV (which can be included following
for instance the statistical procedure of
Ref. \cite{yellin}). Nevertheless our procedure allows us to reproduce
to a good degree of precision the CDMS limit for the standard
isothermal distribution, as shown by the central curve in
Fig. \ref{fig:limits_isothermal}, when this is compared with the upper
bound displayed in Fig. 39 of Ref. \cite{cdms}. Therefore, in our
analysis we adopt our simpler procedure; we have checked that adding a
proper treatment of the efficiency and adopting the statistical
procedure of Ref. \cite{yellin} yields quite similar results.

The upper limits for the isothermal sphere (model A0) are shown in
Fig. \ref{fig:limits_isothermal}, for the three representative values
of $v_0$ and the corresponding choices for the local dark matter
density $\rho_0$, as quoted in Table \ref{tab:intervals}. As already
mentioned, the central curve corresponds to the reference case of
$v_0=220~\mbox{km sec$^{-1}$}$ with $\rho_0=0.3~\mbox{GeV
cm$^{-3}$}$. The upper and lower curves are instead obtained for
$v_0=170~\mbox{km sec$^{-1}$}$ with $\rho_0=0.18~\mbox{GeV cm$^{-3}$}$
and $v_0=270~\mbox{km sec$^{-1}$}$ with $\rho_0=0.71~\mbox{GeV
cm$^{-3}$}$, respectively. An important effect is obviously due to the
different values of $\rho_0$ which are associated to the different
values of $v_0$, as discussed in the previous Section and in
Ref. \cite{bcfs}. However, the difference in the function $\ivmin$ is
quite relevant in the determination of the upper limits, especially at
low WIMP masses.  In order to appreciate the difference in the
exclusion plots, we show in Fig. \ref{fig:ratio} the ratios of the
upper limits obtained with $v_0=170~\mbox{km sec$^{-1}$}$ (lower
curve) and $v_0=270~\mbox{km sec$^{-1}$}$ (upper curve) with respect
to the central $v_0=220~\mbox{km sec$^{-1}$}$ case. The dashed
horizontal lines show the ratios of the corresponding values of
$\rho_0$. We can notice that at low WIMP masses the difference in the
exclusion plots is very large, much larger than the naive ratio of the
corresponding $\rho_0$'s. This is a consequence of the sizable
difference in $\ivmin$ for large $\vmin$, which is the regime relevant
for light WIMPs, as discussed before. The steep behaviour of the
ratios in Fig. \ref{fig:ratio} is a consequence of the fact that the
sensitivity of direct detection to very low WIMP masses (below about
10 GeV) rapidly vanishes.

On the contrary, at large WIMP masses, the difference in the exclusion
plots is much close to what one would expect on the basis of the
difference in the $\rho_0$ values. This is clear from our previous
analysis, since for large WIMP masses the relevant range of $\vmin$ is
in the 100--300 $\mbox{km sec$^{-1}$}$ range, which is where the
difference in $\ivmin$ is small. We can also notice that, for large
WIMP masses, it is even possible to revert the value of the ratio of
the exclusion plots naively obtained by the ratio of the different
$\rho_0$'s: this is a consequence of the behaviour of $\ivmin$ at
small $\vmin$ discussed in the previous Section.

Fig. \ref{fig:limits_isothermal} represents the maximal variability
which occurs for the isothermal sphere: this quantifies the
astrophysical uncertainty connected to this halo model. Confronting
the upper limits with the results obtained for light neutralinos in
supersymmetric models without gaugino--mass universality
\cite{lowneu,lowmassdir,lowind,lowpbar}, we can see that while all the
configuration in the mass range $15\, (8) ~\mbox{GeV} \leq m_\chi \leq
25~\mbox{GeV}$ are excluded for the central (upper) values of $v_0$,
only a small fraction are eliminated when $v_0$ assumes its lower
bound value. Therefore, the conservative attitude which has to be
taken when setting limits makes us to conclude that for the isothermal
sphere, direct detection only mildly constrains the light neutralino
sector of supersymmetric models without gaugino--mass universality
\cite{lowneu,lowmassdir,lowind,lowpbar}, in the 20--40 GeV mass
range. Clearly the variation on the upper limits due to the difference
in the halo properties has consequences also on the exploration of the
supersymemtric parameter space for heavier neutralinos, as is shown in
Fig. \ref{fig:limits_isothermal}. This is relevant also to
gaugino--mass universal models, for which the lower bound on the
neutralino mass exceeds 50 GeV \cite{onehundred}.

The results for the other galactic halo models is shown in
Fig. \ref{fig:limits_distribs}. The differences in the upper limits
can be understood on the basis of the discussion on the isothermal
sphere and on the properties of $\ivmin$ for the different models
presented in the previous Section. We notice that some models, like
C3, D1 and B5 are more constraining, while in the case of models like
A1 and B1 the limits imposed by direct detection are relatively less
severe.

Finally, we report in Fig. \ref{fig:limits_extreme} the summary of our
analysis: together with the standard central isothermal sphere, we
show the more (C3) and less (B1) constraining models we obtain. From
the analysis of this figure, we conservatively conclude that from
direct detection experiments there is currently no constraint on the
light neutralino sector of supersymmetric models without gaugino
universality \cite{lowneu,lowmassdir,lowind,lowpbar}. Should the local
value of the rotational velocity be on its high range (close to
$v_0=270~\mbox{km sec$^{-1}$}$) direct detection could be able to set
stringent limits on these supersymmetric configuration: all the mass
range above 7--8 GeV (depending on the actual halo model) and below 25
GeV would be excluded. Notice that, would this be the case, also the
local density $\rho_0$ would be large (above $0.7~\mbox{GeV
cm$^{-3}$}$, as discussed in Ref. \cite{bcfs} and shown in Table
\ref{tab:intervals}): in this case the neutralino configurations below
7--8 GeV, which are not constrained by direct detection, would be
completely excluded by antiproton searches \cite{lowpbar}. However,
due to astrophysical uncertainties which affect the different
detection rates, currently it is not yet possible to set absolute
limits, neither from indirect detection techniques
\cite{lowind,lowpbar} or, as shown in the present analysis, by direct
detection.

Fig. \ref{fig:limits_extreme} also shows the effect of a lower escape
velocity.  As discussed in the previous Section, this implies a cut in
the high $\vmin$ tail of $\ivmin$: this turns into a weaker
sensitivity of direct detection to low--mass neutralinos. The effect
is especially manifest for the most stringent models, like model C3
with $v_0 =270$ km sec$^{-1}$. For $\vesc=450~\mbox{ km sec$^{-1}$}$,
all neutralino models below 9 GeV are not constrained even for C3
model. For the A0 model with $v_0 =220$ km sec$^{-1}$, there is also a
sizeable difference for light WIMPs, although this is not relevant for
the neutralino configurations. Finally, in the case of model B1 with
$v_0 =170$ km sec$^{-1}$, the lower escape velocity does not produce
differences, since in the high velocity tail $\ivmin$ was already
depressed even for $\vesc=650~\mbox{ km sec$^{-1}$}$, as can be seen
in Fig. \ref{fig:ivmin_distribs}.

For completeness, Fig.\ref{fig:limits_extreme} also shows the upper
limits we obtain, for the isothermal sphere and for the two extreme
cases, for the ZEPLIN detector \cite{zeplin}. The limits we obtain for
the isothermal sphere are slightly higher at low WIMP masses than the
ones quoted by the experimental Collaboration \cite{zeplin}. We trace
this effect to some differences in the analysis of the data (we do not
make use of the ``light response matrix'' discussed in
Ref. \cite{zeplin}, since we do not have it at our
disposal). Fig. \ref{fig:limits_extreme} shows that for very low WIMP
masses ZEPLIN could be slightly more sensitive than
CDMS. Nevertheless, even lowering by a factor of 2 the upper limits we
obtain for ZEPLIN, our conclusions on the limits imposed to light
neutralinos remain unchanged.

Finally, we wish to remind that an annual modulation effect in direct
detection has been observed by the DAMA Collaboration over seven years
\cite{dama}. This result, when interpreted in terms of scalar
WIMP--nucleus interactions, leads to an allowed region in the plane
$\xi\sigman$ vs. $m_\chi$, which extends also to light WIMP masses.
The DAMA Collaboration analysis of Ref. \cite{dama} takes into account
the same variability in galactic halo models of Ref. \cite{bcfs},
which is also used here. It is not possible to make direct comparison
among the DAMA allowed region and the upper limits we obtain here for
CDMS and ZEPLIN, since the DAMA region is the convolution obtained
after varying all the galactic halo models, while the results
presented here refer to single halo model. A proper comparison
between different experimental results can be made {\em only} at a
fixed galactic halo model. Notice that a convolution of our results
would be just the upper curve (model B1) of
Fig.\ref{fig:limits_extreme}.

As for the comparison between the light neutralinos of non--universal
gaugino models and the DAMA allowed region, we comment by reminding
that these light neutralinos are totally compatible with the allowed
DAMA region, as we showed in Ref. \cite{lowmassdir}: they could in
fact explain the annual modulation effect.

\acknowledgments 
We gratefully acknowledge financial support provided by Research
Grants of the Italian Ministero dell'Istruzione, dell'Universit\`a e
della Ricerca (MIUR), of the Universit\`a di Torino and of the
Istituto Nazionale di Fisica Nucleare (INFN) within the {\sl
Astroparticle Physics Project}.


\begin{thebibliography}{99}
\bibitem{note} The class of models discussed in Ref. \cite{bcfs}
represents a quite exhaustive summary of dark matter distributions
functions as possible departures from the isothermal sphere
model. Notice, however, that it does not include the possibility of
non--thermal components in the $f(\vec v)$, which could in principle
modify the high--velocity tail of the velocity DF \cite{helmi}.
\end{thebibliography}
\end{document}